\magnification=1200
\baselineskip15pt
\vglue1in
\hglue 4in
hep-ph/9802xxx

\hglue4in
UM-HE-98-04

\hglue4in
February 1998
\bigskip
\hglue0.9in
{\bf Background Gluon Effects on $\bf B\to X_s \gamma \gamma$}

\vglue0.5in
\hglue1in
S. Rai Choudhury$^{a,b}$ and York-Peng Yao$^a$

\bigskip
\hglue0.2in
a. Randall Laboratory of Physics, University of Michigan, Ann Arbor,

\hglue1.6in
Michigan, 48109 USA
\bigskip
\hglue0.3in
b. Department of Physics, University of Delhi, Delhi 11007, India

\vglue1in
\hglue1.9in
{\bf Abstract}

We consider non-perturbative QCD effects on the energy spectrum of either one of the photons
in $B\to X_s \gamma \gamma$.  These are due to the subprocesses in which a charm quark loop
interacts with a self-consistently produced background static QCD field.  The magnitude is
estimated to be a few percents in $B \to X_s \gamma \gamma$, but can be quite substantial in
$B_ s \to \gamma \gamma$.  An extension of the Euler-Heisenberg Lagrangian is given.

\vglue0.3in
\noindent
PACS:12.38.Bx; 12.38.-t; 13.20.He; 13.40.Hq; 14.65.Fy

\vfill
\eject
\noindent
{\bf I. Introduction}

The flavor changing neutral current inclusive transition $B \to X_s \gamma$ has been a subject of
intense interest during the last few years.  The basic theoretical framework is the Standard electroweak
Model at scales $\sim m_w$ or $m_t$.  QCD short distance corrections [1] are then incorporated in it via
renormalization group technique to yield an effective weak interaction Hamiltonian valid at scales
$\sim m_b$ relevant for B decay processes.  The special role that  $B \to X_s \gamma $ enjoys is
related to its being rather clean and as such is much more model independent compared to exclusive
B-decays.  Its rate has recently been measured [2] with increased accuracy and is in remarkable agreement
with theoretical estimates.  This implies that revelation of New Physics will have to wait and that
further confrontations have to be mounted.  In light of these developments, it is natural to consider
other inclusive channels which as a whole will separate out contributions from various operators in the
effective Hamiltonian.  The branching ratios may be somewhat lower in other novel processes,
but they will still be amenable to
experiments in the new facilities either under construction or being proposed for the not too distant
future.  One such process is the inclusive process $B \to X_s \gamma \gamma$, which is expected to be
$\sim 10^{-2} $ smaller in its branching ratio relative to $B \to X_s \gamma $.  Just like $B \to X_s \gamma$,
however, it is relatively clean after some proper precaution to take out effects
due to strong resonances such as $\eta _c$ at its peak to the two photon spectrum.  It will then provide further
opportunities for testing the
whole technology of weak decays, or better yet in pointing towards some clues of New Physics.

There have been some theoretical activities for the process $B\to X_s \gamma \gamma$, which corresponds
at the quark level to the transition $b \to s \gamma \gamma.$  Calculations were first done on the basis of
pure electroweak theory [3-6] and subsequently improved to include the leading order
renormalization group improved QCD effects [7-9].
These investigations (like most of the investigations for $B \to X_s \gamma $)  are mostly based
on the free quark decay of b, the justification of which is from the heavy quark effective theory (HQET) [10].
According to the argument given, corrections to the free quark results in inclusive processes are
suppressed by powers of $({\Lambda _{QCD}\over m_b})^2$.  Recently, however, Voloshin [11] has shown that corrections
which scale like $({\Lambda _{QCD}\over m_c})^2$ should also exist.  This last number is $\sim 1$ and thus has the
potential of damaging all free quark estimates of inclusive B decays.  For the leading process $B \to X_s
\gamma $, however, it has been shown that the Voloshin type of corrections have small coefficients multiplied to
$({\Lambda _{QCD}\over m_c})^2$ and the overall effects are about $3\%$ of the main term [12-15].  In fact, a more positive
view towards this kind of corrections is to interprete them as systematic accounts of the long distance non-resonant
contributions of $c \bar c$ intermediate states.  We shall subscribe to this constructive point of view and the present article
is an investigation of the related effects in the parallel
process $B \to X_s \gamma \gamma.$  We shall show that while the corresponding corrections apparently have terms which
scale like $({\Lambda _{QCD} \cdot k\over m_c^2})^2$, where k is a typical photon momentum, in addition to
corrections of the $({\Lambda _{QCD}\over m_c})^2$ type, the overall effects are also only a few
percents, just as in $B \to X_s \gamma.$

The plan of this article is as follows: in the next section, we shall briefly summarize some perturbative
results of $b\to s \gamma \gamma $ as a way to introduce our notation.  In section III, we shall discuss the
relevant diagrams which give rise to ${1\over m_c^4}$ and ${1\over m_c^2}$ corrections to this process.  They
come from a charm quark loop, from which an almost static gluon is emitted in addition to the two photons.
Explicit formulaes will be given and matrix elements by HQET will be used to estimate their contributions to
the decay amplitude.  Some numerical work will be presented in the last section, followed by concluding
remarks.

\bigskip
\noindent
{\bf II. $b\to s \gamma \gamma$ in Standard Model with leading QCD corrections}

Radiative $b \to s $ processes is best described in the framework of the following effective Hamiltonian
$$H_{eff}(1) =-{4 G_F \over \sqrt 2}V _{tb}V^\star _{bs} \sum_i C_i O_i, \eqno (1)$$
where $V_{ij}$'s are the CKM matix elements, $G_F$ is the Fermi constant, $C_i$ are the QCD
improved Wilson coefficients, and the $O_i$'s are local operators:
$$O_1=-\bar s_\alpha \gamma ^\mu Lc_\beta \cdot \bar c_\beta \gamma _\mu L b_\alpha,$$
$$O_2=-\bar s_\alpha \gamma ^\mu Lc_\alpha \cdot \bar c_\beta \gamma _\mu L b_\beta,$$
$$O_{3,5}=-\bar s_\alpha \gamma ^\mu Lb_\alpha \cdot \sum_q\bar q_\beta \gamma _\mu (L, R) q_\beta,$$
$$O_{4,6}=-\bar s_\alpha \gamma ^\mu Lb_\beta \cdot \sum_q\bar q_\beta \gamma _\mu (L, R) q_\alpha,$$
$$O_7={e\over 16\pi^2}\bar s_\alpha \sigma ^{\mu \nu} (m_b R + m_sL)b_\alpha \cdot F_{\mu \nu},$$
$$O_8={g_s\over 16\pi^2}\bar s_\alpha \sigma ^{\mu \nu} (m_b R + m_sL)({\lambda ^a\over 2})_{\alpha \beta }b_\beta
\cdot G^a_{\mu \nu}. \eqno (2)$$
\vglue0.5in
\input epsf
\centerline{\epsfxsize=6cm  \epsfbox{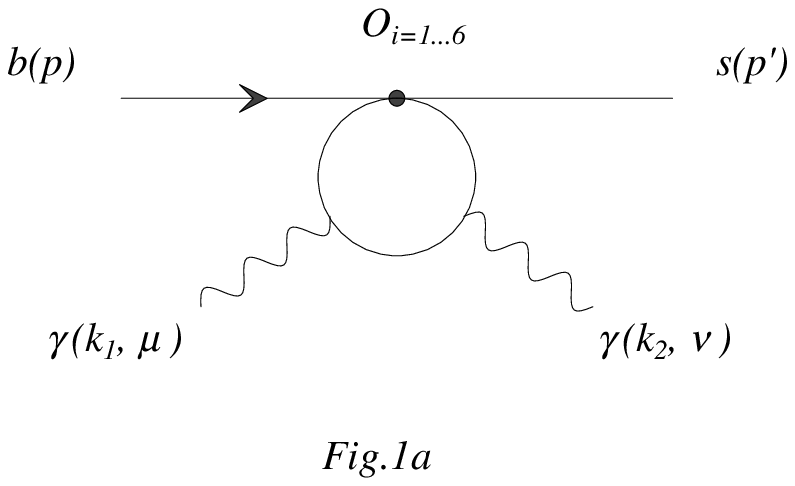} \hfil \epsfxsize=6cm \epsfbox{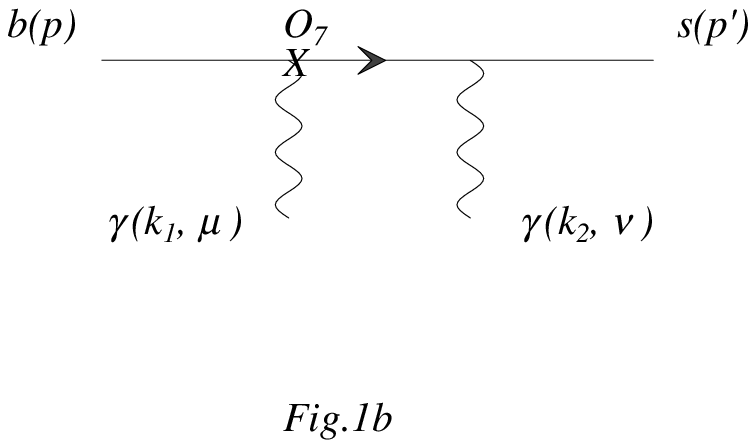} \hfil \epsfxsize=6cm
\epsfbox{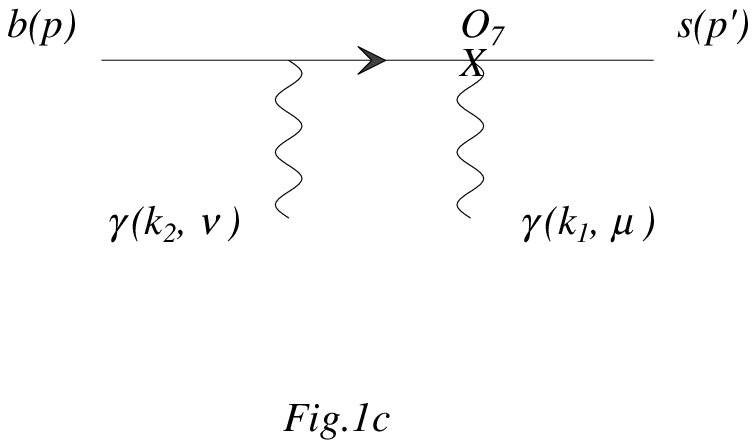}}
{\it Fig.1: One particle irreducible diagram (a) and one particle reducible diagrams (b) and (c) for
$b(p)\to s(p') \gamma (k_1) \gamma(k_2).$  Diagrams with  $(k_1, \ \mu) \leftrightarrow
(k_2, \ \nu) $ should be added for (b) and (c).}
\vglue0.5in

Upon using this effective Hamiltonian, the basic diagrams for the process $b \to s \gamma \gamma $
are shown in fig.(1).  Pure electroweak theory can be reclaimed by retaining only the $C_2 O_2$ term in fig.(1a)
and $C_7 O_7$ term in figs.(1b-1c), with the $C_{2,7}$ given by Inami and Lim [16].  As expected, figs.(1b-1c)
exhibit infra-red divergences, which are cancelled by the virtual radiative corrections to the process
$b \to s \gamma $ as far as any observables are concerned.  Alternatively, a practical measurable rate
for the $b \to s \gamma \gamma $ can be formed by making a cut on the lower end of the energy of either
of the two photons.  This rate in the pure electroweak theory is known to be dominated by the
one particle reducible diagrams of figs.(1b-1c); inclusion of QCD effects enhances their
dominance, because $C_7$ becomes even bigger.  Therefore, in what follows we will assume that the principal
diagrams for $b \to s \gamma \gamma $ are figs.(1b-1c), with only $C_7 O_7$ insertions.  We will compare the
'non-perturbative' gluon corrections to be discussed in the next section relative to this contribution.

The $b \to s \gamma \gamma$ amplitude generated by figs.(1b-1c) strictly speaking cannot be described by an effective
local Hamiltonian.  However, for a fixed value of the invariants $p\cdot k_1, \ p \cdot k_2$ and $k_1 \cdot k_2$,
we can formally write an effective Hamiltonian of the form $\bar b \Gamma _{\mu \nu}s A^\mu A^\nu$, where $\Gamma
_{\mu \nu}$ carries $\gamma$-matrices and momenta.  After summing over the polarizations of the s-quark and the
photons, the transition rate can be represented as  $<B| T(H_{eff}^\dagger (1)H_{eff}(1))|B>$, where the following
transition operator is defined
$$<B|T(H_{eff}^\dagger(1)H_{eff}(1)))|B>=(-{e^2 G_F \lambda _t Q_dC_7\over \sqrt 2 \pi^2})^2
<B|\bar b (W_7^\dagger )^{\mu \nu}{m_s-p'\! \! \! \!/
\over 2 }(W_7)_{\mu \nu}b|B>,\eqno (3) $$
where $\lambda _t=V_{tb}V^\star_{bs}$ and
$$\eqalign {(W_7)_{\mu \nu}={1\over 2}\big[&-{1\over 2p\cdot k_2} k \!\!\! /_1\gamma _\mu (m_b R+m_s L)
(m_b-p\! \! /+  k\! \! \!/_2 )\gamma _\nu \cr & +{1\over 2p'\cdot k_2}\gamma _\nu
(m_s-p\! \! / +k\! \! \! /_1 )k\! \!\!/_1 \gamma _\mu (m_b R +m_s L) \cr
& +(k_1, \ \mu \leftrightarrow k_2 ,\ \nu)\big]. \cr   }\eqno (4)$$
A saliant feature of the spectrum generated by eqs.(3-4) is that as a function of the
invariant mass of the two photons, it peaks at small values. This point will become useful in
simplifying the one-gluon corrections, which we now immediately turn to.

\bigskip
\noindent
{\bf III. ${1\over m_c^2}$ and ${1\over m_c^4}$ corrections to $b \to s \gamma \gamma$}

Because the coefficient $C_2$ is much bigger than $C_{4,6}$, we shall discard effects due to
$O_{4,6}$ in our discussion.  As for the operator $O_2$,  we rewrite it by a Fierz transformation as
$$ O_2 ={1\over 3}  O_1+2\tilde O_1, \eqno (5)$$
where
$$\tilde O_1 =-\bar c \gamma _\mu {\lambda ^a \over 2}Lc \cdot \bar s \gamma ^\mu {\lambda ^a \over 2}L b.  \eqno (6)$$
\vglue0.5in
\input epsf
\centerline{\epsfxsize=5cm  \epsfbox{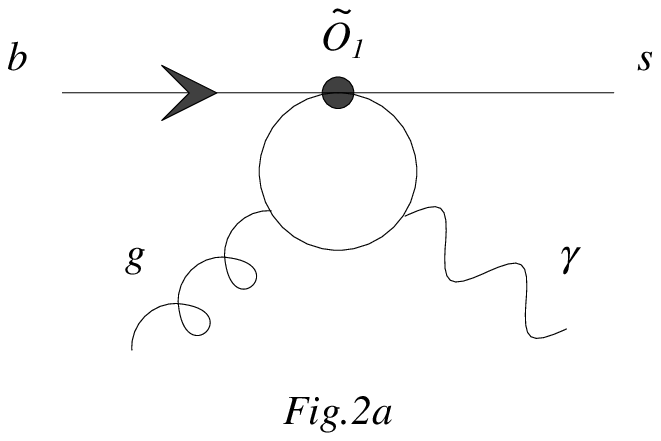} \hfil \epsfxsize=5cm \epsfbox{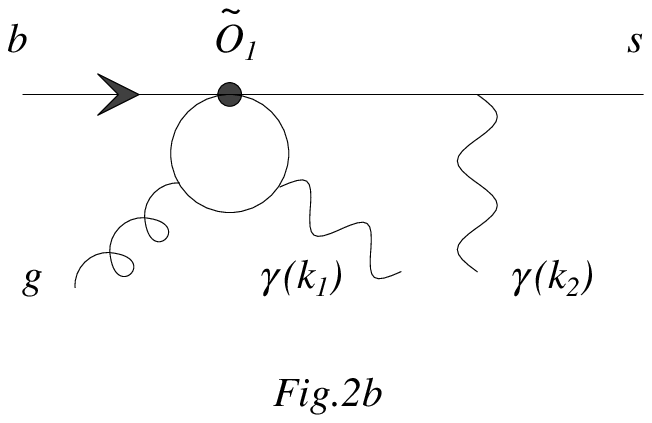} \hfil \epsfxsize=5cm
\epsfbox{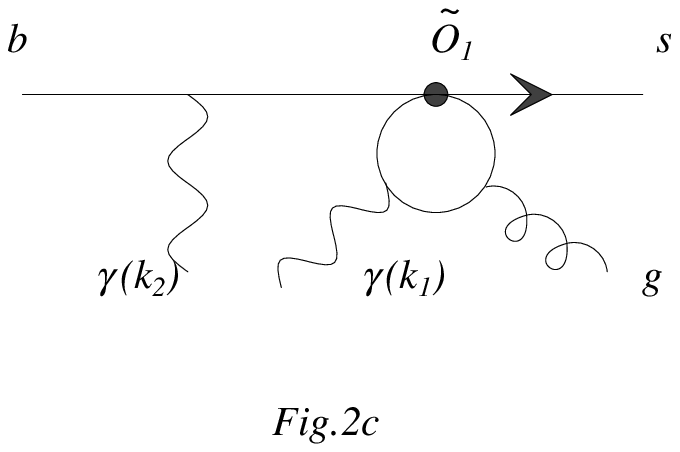}}
{ \it Fig.2: (a) A charm quark loop which generates an effective $bs\gamma g$ vertex; (b) and (c) represent
one particle reducible diagramss, which involve the $bs\gamma g $ vertex above.  Diagrams with
$(k_1, \ \mu) \leftrightarrow (k_2, \ \nu) $ should be added.}
\vglue0.5in
The operator $\tilde O_1$ generates a b-s transition with a c quark loop.  Hooking onto it a photon and a
gluon, fig.(2a) leads to an effective $bs \gamma g$ Voloshin local effective vertex
$$O_2 \ \ \ \to \ \ \ O_{Vol}={-eg_s\over 72 \pi^2 m_c^2}\bar s \gamma _\rho
{\lambda ^a \over 2} L b G^a _{\alpha \beta}\partial ^\beta \tilde F^{\rho \alpha}, \eqno (7)$$
in the soft gluon limit, where $\tilde F^{\rho \alpha}={1\over 2}\epsilon ^{\rho \alpha \lambda \kappa }
F_{\lambda \kappa}, \ \epsilon ^{0123}=1)$ is the dual tensor to the electromagnetic field,
and $G^a_{\alpha \beta} $ is the gluon field tensor.  In this approximation which was used successfully and
justified in ref.[11-15], the gluon inside the
B-meson has been treated as a static field.  The effective $bs \gamma g$-vertex in eq.(7) generates
IPR $b \to s \gamma \gamma g$ amplitudes as shown in figs.(2b-2c).  They cannot strictly be written as local
effective interactions.  However, as in the case of figs.(1b-1c) discussed before, for a given set of values
for the momentum variables, the matrix element of figs.(2b-2c) (plus those with $k_1, \mu \leftrightarrow k_2,  \nu$)
can be formally looked upon as arising out of a Hamiltonian
$$H_{eff}(2)=\big(-{4G_F\over \sqrt 2}\lambda _t\big)\big(-{e^2Q_d C_2\over 72 \pi^2m_c^2}\big) O_R ,\eqno (8)$$
with ($G_{\alpha \beta}\equiv g_s G_{\alpha \beta}^a {\lambda ^a\over 2}$)

$$O_R=\bar s \Gamma ^{\mu \nu }_{\alpha \beta} G^{\alpha \beta}bA_\mu (k_1)A_\nu (k_2),
\eqno (9)$$
and
$$\eqalign {\Gamma ^{\mu \nu }_{\alpha \beta }=&{1\over 2p'\cdot k_2}\gamma ^\nu (m_s-p\! \! /+k\!\!\!/_1)
\epsilon ^\mu_{\lambda \rho \alpha}k_1^\lambda \gamma ^\rho Lk_{1 \beta}
-{1\over 2p \cdot k_2}\epsilon ^\mu_{\lambda \rho \alpha}  k_1^\lambda \gamma ^\rho Lk_{1 \beta}
(m_b-p\! \! / +k\! \! \! /_2 )\gamma ^\nu \cr & +(k_1, \ \mu \leftrightarrow k_2, \nu).}  \eqno (10)$$
\vglue0.5in
\input epsf
\centerline{\epsfxsize=7cm  \epsfbox{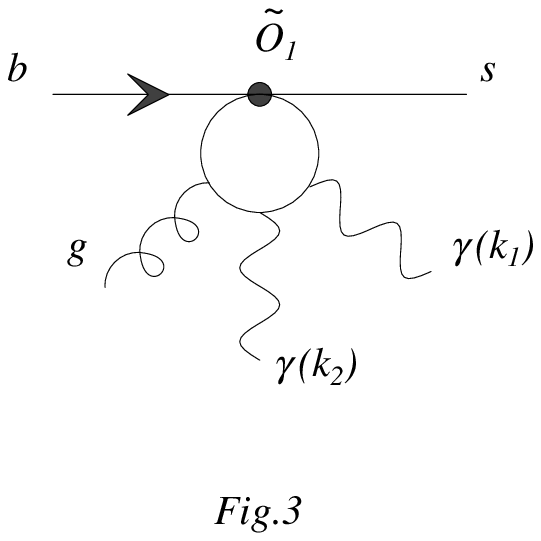}}
{ \it Fig.3: A one particle irreducible diagram due to a charm quark loop, which generates an effective
$b \to s \gamma \gamma g$ vertex.  Five other diagrams due to permutations of $\gamma $'s and g should
be added. }
\vglue0.5in

Finally, fig.(3) represents an irreducible diagram for $b \to s \gamma \gamma g$ transition generated by the
charm loop, again via $\tilde O_1$ of eq.(6), which scales like $1\over m_c^4$.  Unlike the vertex in
fig.(2a), only the vector part of $\tilde O_1$ now gives a non-vanishing contribution to it.  Fig.(3) can be
written as an effective Hamiltonian, which is basically the photon-photon scattering amplitude.  The general
expression for arbitrary $k_1, \ k_2$ and $k_3$ is quite complicated [17], but, in the same spirit taken earlier, the
gluon inside the B-meson can be treated as a background static field.    We
first form the gluon field tensor and then consider the limit with its momentum $k_3 \to 0$.
In this provision, we obtain

$$\eqalign {H_{eff}(3)= &\big({Q_u^2e^2g_sC_2\over 16\pi^2}\big) (-{4G_F\lambda_t\over \sqrt 2})\cr & \times
\big(  \ \ G^{\delta \kappa}_a F_{\kappa \mu}F^{\mu \nu}H_{\nu \delta}^a \big[(-{112\over t^2}-{2\over m_c^2t})
+(64{m_c^2\over t^2}-{8\over t})I^1_{00} \cr
&  \ \ \ \ \ \ \ \ \ \ \ \ \ \ \ \ \ \ \ \ \ \ \ +(1280{m_c^4\over t^4}+672{m_c^2\over t^3} +{88\over t^2})fln\big] \cr
&+G^{\delta \kappa}_aF_{\kappa \mu}H^{\mu \nu}_aF_{\nu \delta}
\big[ {40\over t^2}-16{m_c^2\over t^2}I^1_{00} \cr
&\ \ \ \ \ \ \ \ \ \ \ \ \ \ \ \ \ \ \ \ \ \ \ +(-512 {m^4\over t^4}-224{m_c^2\over t^3}-{24 \over t^2})fln\big]\cr
&+ G^{\delta \kappa}_aF_{\delta \kappa}F^{\mu \nu}H_{\mu \nu }^a \big[ ({32\over t^2}+{1\over 3m_c^2t^2})
+(-16 {m_c^2\over t^2}+{4\over t})I^1_{00}\cr
& \ \ \ \ \ \ \ \ \ \ \ \ \ \ \ \ \ \ \ \ \ \ \ + (-384 {m^4\over t^4}-208 {m^2\over t^3}-{28\over t^2})fln\big] \cr
& + G^{\delta \kappa}_a H^a_{\delta \kappa}F^{\mu \nu}F_{\mu \nu} \big[ ({11\over t^2}+
{1\over 2m_c^2t})-8{m_c^2\over t^2}I^1_{00}  \cr
& \ \ \ \ \ \ \ \ \ \ \ \ \ \ \ \ \ \ \ \ \ \ \ +(-112{m_c^4\over t^4}-60{m_c^2\over t^3}-{8\over t^2})fln\big]\cr
&  -(\partial _\lambda \partial ^\alpha F^{\mu \nu})(\partial _\delta F_{\mu \nu}) H_\alpha ^a G^{\delta \lambda}_a
\big[{192\over t^3}+(-96{m_c^2\over t^3}+ {16\over t^2})I^1_{00}\cr
& \ \ \ \ \ \ \ \ \ \ \ \ \ \ \ \ \ \ \ \ \ \ \
+(-2304 {m_c^4\over t^5}-1152 {m_c^2\over t^4}-{144\over t^3}) fln\big] \big),\cr  }\eqno (11)$$
where $$I^1_{00}=\int _0^1 {dx_1dx_2dx_3 \delta (1-x_1-x_2-x_3)\over m_c^2+x_1x_2t}
={1\over 2t}ln^2({\sqrt{4m_c^2+t}+\sqrt t\over \sqrt{4m_c^2+t}-\sqrt t}), $$
$$fln=({t\over t+4m_c^2})^{3/2}ln({\sqrt {t\over m_c^2}}{1+\sqrt{1+{4m_c^2\over t }}\over 2}),$$
$$H_\mu^a=\bar s \gamma _\mu L{\lambda^a\over 2}b , \ \  \ \
H_{\mu \nu}=\partial _\mu H_\nu ^a-\partial _\nu H_\mu^a. \eqno (12)$$
have been defined for $t\equiv 2k_1\cdot k_2 >0$.  The prescription $m_c^2-i\epsilon$ should be taken
to analytically continue to the physical region $t<0$.
As we mentioned in the last section, the region of interest for the spectrum
$b \to s \gamma \gamma $ turns out to be when one of the momenta of the two photons
becomes soft.  Then the expression in eq.(11) is further simplified into a local one

$$\eqalign {H_{eff}(3)= \big({Q_u^2e^2g_sC_2\over 16\pi^2m_c^4}\big) (-{4G_F\lambda_t\over \sqrt 2})
&\big[  \ \ G^{\delta \kappa}_a F_{\kappa \mu}F^{\mu \nu}H_{\nu \delta}^a (-{14\over 45})
+ G^{\delta \kappa}_aF_{\kappa \mu}H^{\mu \nu}_aF_{\nu \delta}(-{7\over 45})\cr
& + G^{\delta \kappa}_aF_{\delta \kappa}F^{\mu \nu}H_{\mu \nu }^a({1\over 9})
+ G^{\delta \kappa}_a H^a_{\delta \kappa}F^{\mu \nu}F_{\mu \nu} ({1\over 18}) \big]
,\cr  }\eqno (13)$$
which is in momentum space
$$H_{eff}(3)= \big(-i{Q_u^2e^2C_2\over 16\pi^2m_c^4}\big) (-{4G_F\lambda _t\over \sqrt 2})
\bar s (O_{IR})^{\delta \kappa}_{\mu \nu}G_{\delta \kappa }b A^\mu(k_1)A^\nu(k_2),\eqno (14)$$
where
$$\eqalign {(O_{IR})^{\delta \kappa}_{\mu \nu}=&
{1\over 45}
\times \big[ 3k_1^\nu k_2^\mu ((k_1+k_2)^\delta \gamma^\kappa -(k_1+k_2)^\kappa \gamma^\delta)
+14(k_2^\mu \gamma^\nu -k_1^\nu \gamma^\mu)(k_1^\delta k_2^\kappa-k_1^\kappa k_2^\delta) \cr
& +7g^{\mu \nu}(k_1^\delta k_2^\kappa-k_1^\kappa k_2^\delta)(k_1-k_2)\cdot \gamma
+3g^{\mu \nu}((k_1+k_2)^\kappa \gamma^\delta -(k_1+k_2)^\delta \gamma^\kappa)k_1 \cdot k_2 \cr
&-7(g^{\mu \delta}g^{\nu \kappa}-g^{\mu \kappa}g^{\nu \delta}) k_1 \cdot k_2(k_1-k_2)\cdot \gamma\cr
& +3(g^{\mu \kappa} k_1^\delta -g^{\mu \delta}k_1^\kappa)k_1^\nu k_2 \cdot \gamma
+ 3(g^{\nu \kappa} k_2^\delta -g^{\nu \delta}k_2^\kappa)k_2^\mu k_1 \cdot \gamma  \cr
& +3(g^{\mu \delta}k_1^\kappa-g^{\mu \kappa}k_1^\delta)\gamma^\nu k_1\cdot k_2
+  3(g^{\nu \delta}k_2^\kappa-g^{\nu \kappa}k_2^\delta)\gamma^\mu k_1\cdot k_2    \cr
& +7(g^{\mu \delta} k_2^\kappa  - g^{\mu \kappa}k_2^\delta )k_1^\nu (k_1+k_2)\cdot \gamma
+ 7(g^{\nu \delta} k_1^\kappa  - g^{\nu \kappa}k_1^\delta )k_2^\mu (k_1+k_2)\cdot \gamma  \cr
& + 14(g^{\mu \kappa}k_2^\delta -g^{\mu \delta }k_2^\kappa)\gamma^\nu k_1\cdot k_2
+  14(g^{\nu \kappa}k_1^\delta -g^{\nu \delta }k_1^\kappa)\gamma^\mu k_1\cdot k_2\big]L . \cr } \eqno(15) $$

The quarks inside the B-meson are in constant interaction with each other via soft gluon exchanges.
The $B \to X_s \gamma \gamma$ transition can thus be regarded as the quark transition $b \to s
\gamma \gamma $ in the presence of a background gluon field.  This is the approach of Voloshin, who further regarded
this background gluon field as static.  To first order in the strong QCD coupling, the effective Hamiltonians
$H_{eff}(2)$ of eq.(8) amd $H_{eff}(3)$ of eq.(14), in which a zero momentum gluon is emitted from a c-quark
loop, would then implement this dynamics.  The corresponding amplitudes so generated would add coherently with the
main amplitude of eq.(3).  These correction terms are small (to be justified a posteriori), so that the addition
to the principal transition operator would be the interference terms of amplitudes of fig.(2-3) with those of
figs.(1b-c).

The interference terms between $H_{eff}(1)$ due to $C_7O_7$ and $H_{eff}(2)$ are
$$\eqalign {<B|T(H^\dagger_{eff}(1)H_{eff}(2))|B>= &(-{e^2Q_d G_F \lambda _t C_7\over \sqrt 2 \pi^2})
(-{4G_F\lambda _t\over \sqrt 2})
(-{e^2Q_dC_2\over 72\pi^2m_c^2}) \cr
& <B|\bar b (W^\dagger _7)_{\mu \nu}{m_s-p' \!\! \! \! /\over 2}\Gamma^{\mu \nu}_{\alpha \beta}
G^{\alpha \beta}b|B>,\cr  } \eqno (16)$$
and the complex conjugate.  Corresponding to the interference of $H_{eff}(1)$ and $H_{eff}(3)$, we have
$$\eqalign {<B|T(H^\dagger _{eff}(1)H_{eff}(3))|B>=& (-{e^2 Q_dG_F\lambda _t C_7\over\sqrt 2 \pi^2})
({-iQ_u^2 e^2C_2\over 16\pi^2 m_c^4})({-4G_F \lambda _t \over \sqrt 2 })\cr
& <B|\bar b (W_7^\dagger)_{\mu \nu}{m_s-p'\!\! \! \! /\over 2}(O_{IR})^{\mu \nu}_{\delta \kappa}
G^{\delta \kappa}b|B>, \cr    }\eqno (17)$$
and its complex conjugate.  Eqs.(3, 16 and 17) give the total transition amplitude
$$\eqalign {<T>=&<B|T(H_{eff}^\dagger (1)H_{eff}(1))|B>+\big(<B|T(H_{eff}^\dagger (1)H_{eff}(2))|B>+c.c.\big)\cr
&+\big(<B|T(H_{eff}^\dagger (1)H_{eff}(3))|B>+c.c.\big) \cr}\eqno(18)$$

We rely on HQFT to evaluate the matrix elements [18]
$$<B(v)|\bar b \Gamma b|B(v)>={1\over 2}Tr\big[{1-v \! \! \! / \over 2}\Gamma\big], \eqno (19)$$
and
$$<B(v)|\bar b \Gamma G_{\alpha \beta}b|B(v)>={\lambda_2 \over 2}
Tr\big[{1-v \! \! \! / \over 2}\Gamma {1-v \! \! \! / \over 2}\sigma_{\alpha \beta}\big], \eqno (20)$$
where $v$ is related to the momentum of b by $p=m_bv$, $\Gamma $ is any Dirac structure, and $\lambda _2$
is related to the $B^\star-B$ mass splitting with a numerical value $\lambda _2=0.12 (Gev)^2$.
We have normalized $<B(v)|B(v)>=1.$  The rate for  $b(p)\to s(p')\gamma (k_1) \gamma (k_2)$ is
$$d\Gamma={1\over 2(2\pi)^5}\delta^4(p-p'-k_1-k_2){d^3p'\over (p')^0}{d^3k_1\over 2\omega_1}
{d^3k_2\over 2\omega_2}<T>.\eqno (21)$$

\bigskip
\noindent
{\bf IV. Numerical Results and Discussion}
\bigskip
We would like to estimate the relative contributions of the three terms in eq.(18).  As discussed earlier,
we shall interprete corrections from the second and the third term as an indication of the long distance
non-perturbative effects away from the peaks of the appropriate $c \bar c$ resonances. In this regard, the
case for $\eta_c$  was addressed by the authors in ref.(8).  They concluded that its effects on the spectrum
of the two photon invariant mass are very localised.  Our study here is on a different aspect of the same
kind of issues.

We find it convenient to present our numerical work in a quantity introduced by these authors.  This is
the spectrum of the photon with the lower energy and that of the photon with the higher energy, defined by
$${d\Gamma ^{L,H}\over dk_1}=\int {d\Gamma \over dk_1dk_2}\theta (\pm k_2 \mp k_1)dk_2,\eqno (22)$$
where the integration domain, as in ref.(8), is restricted by the requirements that the energy of each
photon be larger than $E_\gamma ^{min}=100 Mev$, and that the angles between any two outgoing particles be
bigger than 20 degrees.  These make it experimentally practical to distinguish from $b \to s \gamma$.
We fix $\mu=m_b=4.8 Gev, \ m_c=1.5 Gev$ and $m_s=450 Mev$, which lead to $C_2=1.09$ and $C_7=-0.31$.
Figs.(4-5) exhibit $50\times $ and $100\times$ the two interference terms separately vs. the main term of
the spectrum from Eq.(18).  Note that the photon energy $k_1$ is measured in unit of $m_b$.  Also, all the curves
have been normalized to ${d\Gamma ^L\over dk_1}$ of the main term at $k_1=0.086$.
The results show that despite the ${1\over m_c^2}$ or ${1\over m_c^4}$ scaling the static
gluon corrections from the charm loop is at most a few percents and hence not experimentally observable
at this time.  This is similar to the results for the one photon process, where however the corrections
scale only as ${1\over m_c^2}$.

\vglue0.5in
\input epsf
\centerline{\epsfxsize=6cm  \epsfbox{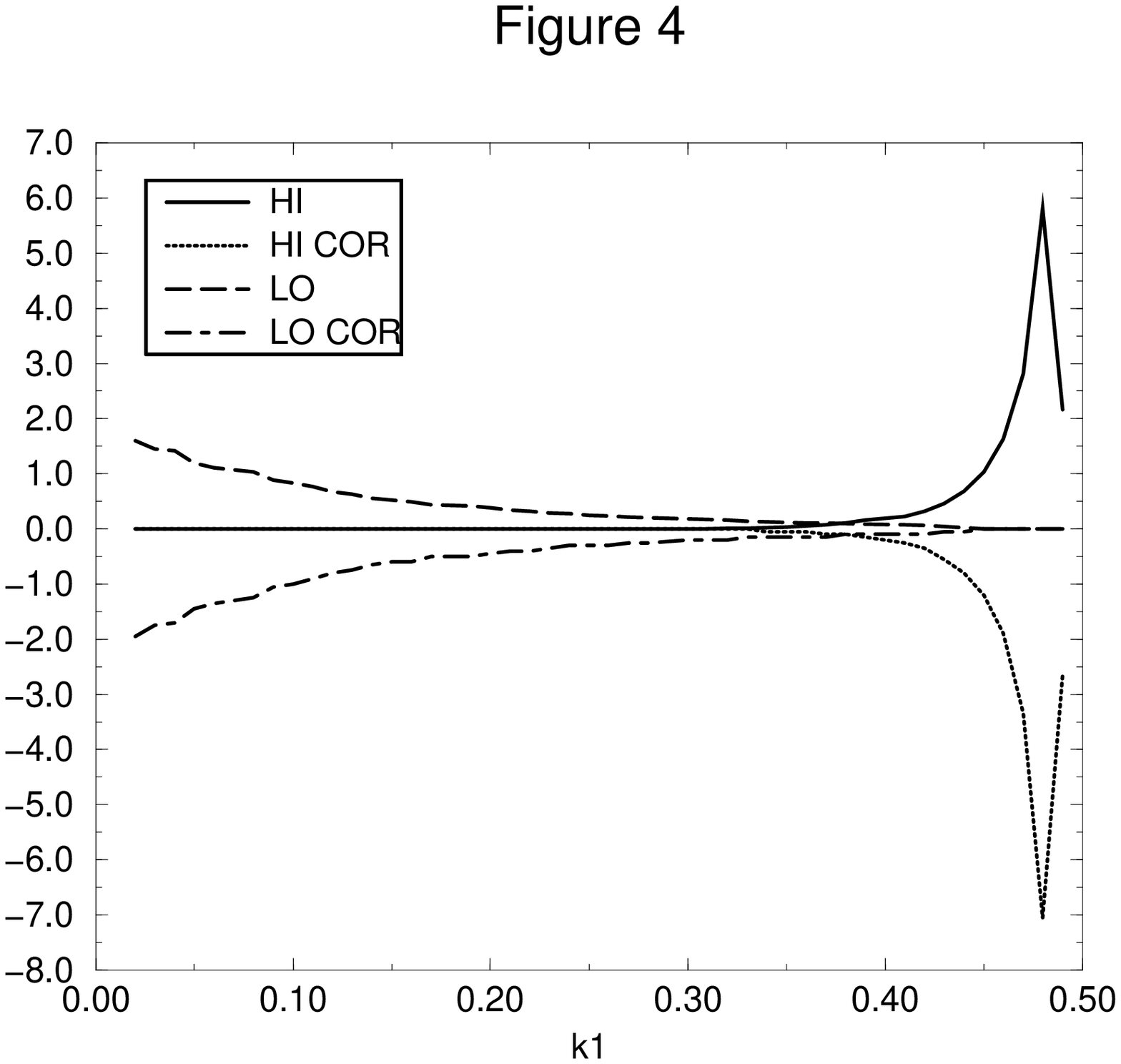} \hfil \epsfxsize=6cm \epsfbox{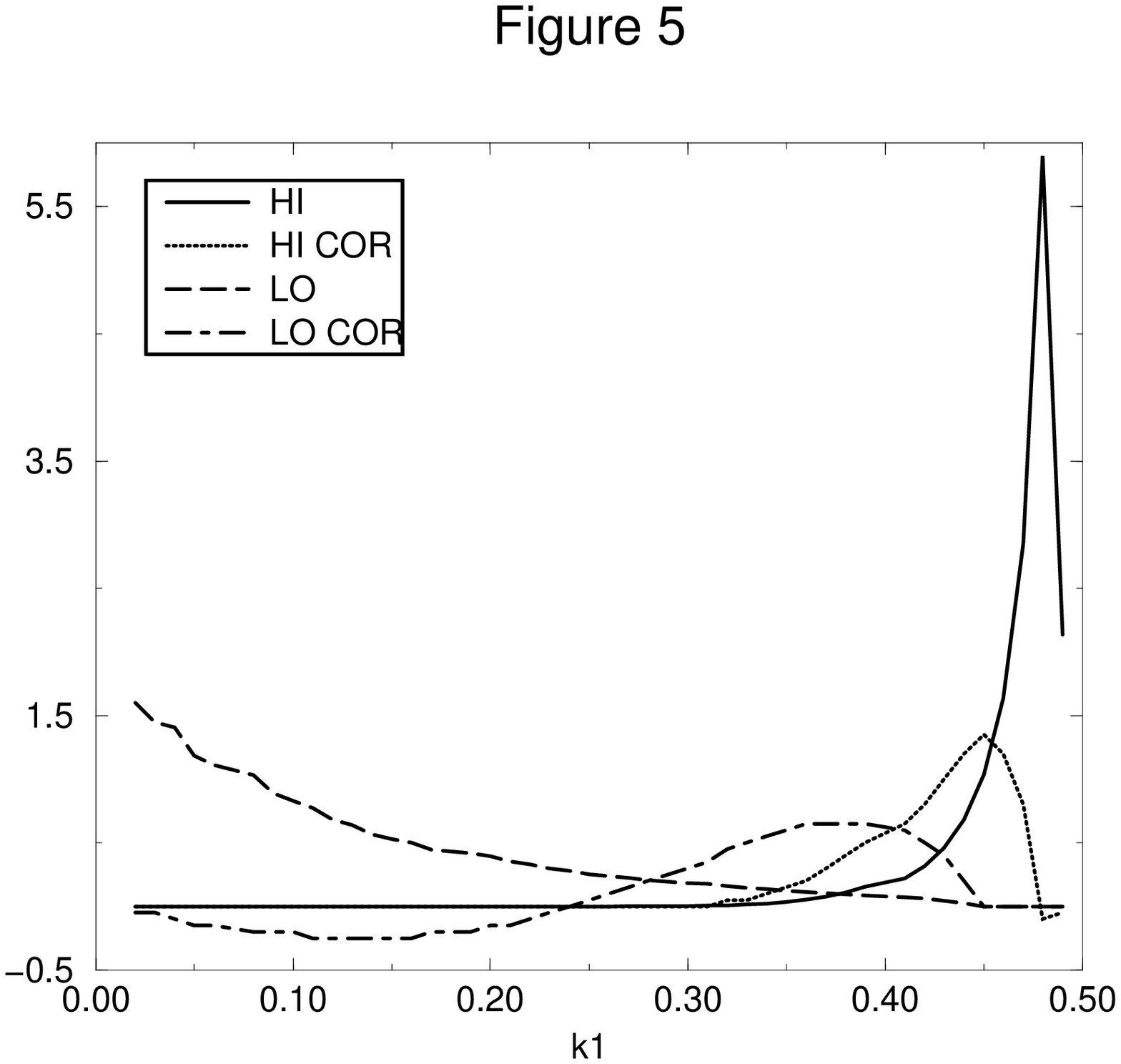} }
{\it Fig.4: The spectra of higher energy photon ${d\Gamma ^H\over dk_1}$ (HI) and lower energy photon
${d\Gamma ^L\over dk_1}$ (LO) due to the first main term of Eq.(18) and $50\times$ of the corresponding
quantities due to the second term, denoted as HICOR and LOCOR, respectively, are shown. $ k_1$ is measured
in unit of $m_b$.  All curves have been devided by the value of LO at $k_1=0.086$.

\bigskip

Fig.5: The spectra of higher energy photon ${d\Gamma ^H\over dk_1}$ (HI) and lower energy photon
${d\Gamma ^L\over dk_1}$ (LO) due to the first main term of Eq.(18) and $100\times$ of the corresponding
quantities due to the third term, denoted as HICOR and LOCOR, respectively, are shown. $ k_1$ is measured
in unit of $m_b$.  All curves have been devided by the value of LO at $k_1=0.086$.}
\vglue0.5in
The fact that a correction term which apparently goes like $({k_1\cdot k_2\over m_c^2})^2$, where k's are the
typical photon momenta, is small is obviously because the two photon invariant mass stays small in the regions
with appreciable rate.  This situation would change dramatically for the exclusive process
$B_s \to \gamma \gamma $ where the invariant mass $-2k_1 \cdot k_2 =m_b^2$.  One can expect contributions
from fig.(3) to be of order $({m_b\over m_c})^4$.  On the other hand, $B_s \to \gamma \gamma $ is an
exclusive channel, and therefore the same technique applied here and in refs.[11-15] cannot be directly
relied upon.  We are presently engaged in evaluating these effects.

Before closing, we would like to point out that eq.(13) is in agreement with the Euler-Heisenberg
Lagrangian [19] and that eq.(11) is an exact extension of the photon-phonton scattering amplitude to a
situation when one of the photons ($G_{\mu \nu}^a$ in this case) is static while two on-shell
photons can have any four momenta.  The last photon is off-shell to give non-trivial kinematics.

\bigskip
\noindent
{\bf Acknowledgements}

SRC would like to thank the Department of Physics and the particle theory group of the University
of Michigan for hospitality.  The work by YPY has been partially supported by the U. S. Department
of Energy.
\vfill
\eject

\bigskip
\noindent
{\bf References}

\noindent
[1] B. Grinstein, R. Springer, and M. B. Wise, Phys. Lett. ${\bf B202}, 138$ (1988); Nucl. Phys.

\hglue0.1in ${\bf B339}$, 269 (1990);

\noindent
R. Grigjanis, P. J. O. Donnel, M. Sutherland, and H. Navelet, Phys. Lett. ${\bf B 213}$, 355

\hglue0.1in (1998); ${\bf B 286}$, 413 (E) (1992);

\noindent
M. Misiak, Phys. Lett. ${\bf B 269}$, 161 (1991) ; Nucl. Phs. ${\bf B 393}$, 23 (1993); ${\bf B 439}$, 461 (E)

\hglue0.1in (1995);

\noindent
K. Adel and Y.-P. Yao, Mod. Phys. Lett. ${\bf A 8}$, 1679 (1993); Pys. Rev. ${\bf D 49}$, 4945 (1994);

\noindent
M. Chiuchini, E. Franco, G. Martinelli, L. Reina, and L. Silvestrini, Phys. Lett. ${\bf B 316}$,

\hglue0.1in 127 (1993); Nucl. Phys. ${\bf B 421}$, 41 (1994);

\noindent
G. Cella, G. Curci, G. Ricciardi, and A. Vicer${\acute e}$, Phys. Lett. ${\bf B 325}$, 227 (1994);
Nucl.

\hglue0.1in Phys. ${\bf B 431}$, 417 (1994);

\noindent
K. G. Chetyrkin, M. Misiak, and M. M\"unz, Phys. Lett. ${\bf B 400}$, 206 (1997);

\noindent
C. Greub and  T. Hurth, Phys. Rev. ${\bf D 56}$, 2934 (1997);

\noindent
A. Buras, A. Kwiatkowski, and N. Pott, hep-ph/9710336.

\noindent
[2] M. S. Alam et al., Phys. Rev. Lett. ${\bf 74}$, 2885 (1995).

\noindent
[3] G.-L.Lin, J. Liu, and Y.-P. Yao, Phys. Rev. Lett. ${\bf 64}$, 1498 (1990); Phys. Rev. ${\bf D 42}$,

\hglue0.1in 2314 (1990).

\noindent
[4] H. Simma and D. Wyler, Nucl. Phys. ${\bf B 344}$, 283 (1990).

\noindent
[5] S. Herrlich and J. Kalinowski, Nucl. Phys. ${\bf B 381}$,501 (1992).

\noindent
[6] L. Reina, G. Ricciardi and A. Soni, Phys. Lett. ${\bf B 396}$, 231 (1997).

\noindent
[7] C.-H. V. Chang, G.-L. Lin, and Y.-P. Yao, Phys. Lett. ${\bf B 415}$, 395 (1997).

\noindent
[8]  L. Reina, G. Ricciardi, and A. Soni, Phys. Rev. ${\bf D 56}$, 5805 (1997).

\noindent
[9] G. Hiller and E. O. Iltan, hep-ph/9704385.

\noindent
[10] J. Chay et al. Phys. Lett. ${\bf B 247}$, 399 (1990); A. F. Falk, M. Luke, and M. J. Savage,

\hglue0.1in Phys. Rev. ${\bf D 49}$, 3367 (1994).

\noindent
[11] M. B. Voloshin, Phys. Lett. ${\bf B 397}$, 275 (1997).

\noindent
[12] A. K. Grant, A. G. Morgan, S. Nussinov, and  R. D. Peccei, Phys. Rev. ${\bf D 56}$, 3151

\hglue0.1in (1997).

\noindent
[13] Z. Legeti, L. Randall, and M. B. Wise, Phys. Lett. ${\bf B 402}$, 178 (1997).

\noindent
[14] A. Khodjamirian, R. R\"ukle, G. Stoll, and D. Wyler, Phys. Lett. ${\bf B 402}$, 167 (1997).

\noindent
[15] G. Buchalla, G. Isidori, and S.-J. Rey, hep-ph/9705253.

\noindent
[16] T. Inami and  C. S. Lin, Prog. Theor. Phys. ${\bf 65}$, 297 (1981); ${\bf 65}$, 1772 (1981).

\noindent
[17] R. Karplus and M. Neuman, Phys. Rev. ${\bf 80}$, 380 (1950).

\noindent
[18] For a review, see M. Neubert, Phys. Reports ${\bf 245}$, 260 (1994).

\vfill \eject

\noindent
[19] H. Euler, Ann. Phys. (Leipzig) ${\bf 26}$, 398 (1936); W. Heisenberg and H. Euler, Zeit.

\hglue0.1in Phys. ${\bf 98}$, 714 (1936).
\vfill
\eject
\noindent

\end

{\bf Figure Captions}
\bigskip
\noindent

\bigskip
\noindent

\bigskip
\noindent

\bigskip
\noindent

\end